\documentclass[11pt,twoside]{article}
\usepackage{asp2010}
\resetcounters
\begin{document}
\title{The Future of ADASS}
\author{Nuria~P.~F.~Lorente and Keith~Shortridge
\affil{$^1$Australian Astronomical Observatory, PO Box 915, North Ryde, NSW 1670, Australia}
}

\begin{abstract}
ADASS has been a successful conference series for 24 years. If it is to continue to be successful
and relevant we need to ensure that it provides what we as a community need from an annual
conference.
Earlier this year the ADASS Program Organising Committee conducted a survey on the content, style
and governance of ADASS, in order to ascertain the conference needs of our community of astronomy
software, methods and algorithms providers and users. 140 people participated in the survey:
familiar faces, newcomers and a significant number of people who have yet to attend an ADASS.

We summarise the Birds of a Feather session held on 7 October 2014, which discussed the findings of
the survey and the shape that the community would like future ADASS meetings to take: What do we
like of the current format? What would we change? What can we do to make ADASS fit our current and
future needs?
If we are to ensure that ADASS is vibrant, interesting and at the cutting edge of our subject we
need to take collective responsibility for shaping its future.
\end{abstract}

\section{Introduction}
ADASS is the major annual conference for those interested in astronomy
software and systems, from architects and developers to users and
managers.
Its 24 year history indicates its usefulness to individuals and
institutions, as a place to showcase our latest work, to learn new
techniques and discover which have been successfully used in our
specialist field; to discuss emerging technologies with colleagues
with similar interests and generally to work towards continually
improving the overall shape of the field of astronomical software.

Success is no reason for complacency, however. Over the last two
decades we have seen the needs of our community change, and indeed our
demographic has itself changed. On the broader astronomical stage
we're seeing our travel budgets shrinking, and many find themselves in
the uncomfortable position of having to decide whether their single
overseas conference for the year will focus on their science or
on the technology aspects of their work
If it is to continue being a useful and worthwhile conference, ADASS
must adapt, keeping up with the needs of the ADASS community and the
astronomical community as a whole, while keeping in mind our changing
demographics.

To this end the ADASS Program Organising Committee (POC) ran a survey
in early 2014 to ascertain the needs of the community and the
direction and shape which future ADASS meetings should take. 140
people participated (as a comparison, a typical ADASS has around 270
attendees), 84\% of whom had attended 2 or more ADASS conferences,
12\% had attended 1 and 14\% had never been to an ADASS.

\section{Results of the 2014 Survey and the BoF Discussion}
\begin{description}
\item[Identity] We advertise the ADASS conference as {\em``a forum for
{\bf \em scientists}, {\bf \em developers} and {\bf \em programmers} working in areas related to algorithms,
software and systems for the acquisition, reduction, analysis, and dissemination
of astronomical data''}.
The 140 survey participants identified themselves as astronomers
(69\%), software engineers (38\%), programmers (30\%), software
architects (29\%), managers (28\%) and computer scientists (22\%),
with a small number of instrument scientists, and database, archive
and data specialists\footnote{Participants were able to choose
  multiple categories, hence the sum of all the categories is greater
  than 100\%.}.
The fact that 31\% of people did not self-identify as astronomers shows that we
have moved from a community of ``astronomers who code'', to a more
diverse one which includes people whose primary expertise is in
engineering. This reflects the trend we see in our institutes and
telescope facilities, and it is important that the focus and program
of the ADASS conference also evolves over time to keep up with the
changing needs of our community.
The very small numbers of postdocs (6) and students (4) who
participated in the survey is compatible with the small number from
these two groups who attend the conference. This was discussed in the
BoF and it was agreed that mechanisms should be put in place to
encourage more students and early career postdocs to come to ADASS.

\item[Themes] Each year's conference is based around a handful of key
  themes. These are chosen by the POC and are intended to reflect the
  current work, trends and needs of our community, including
  suggestions made by the conference attendees\footnote{A list of the
    themes chosen for the last several conferences can be found at
    \url{http://adass.org/docs/ADASSThemeHistory.pdf}}.
The BoF participants agreed that final responsibility for theme selection
should rest with the POC, and also that the community should be
actively encouraged to nominate key themes. The Appgree mobile
application\footnote{\url{http://www.appgree.com}}, which was tested during
this conference, was put forward as a possible way of facilitating participation
and of canvassing community preferences on the submitted topics.
For ADASS XXV the POC plans to use Appgree to receive theme suggestions
and hopes to see wide participation from the community.
Additionally, 88 people submitted key topic suggestions through the survey, and these will
also inform the selection of key themes for the next several conferences.

\item[Conference Content]
Both the survey and the BoF discussion agreed that there is a need
for greater focus on algorithms, statistics, analysis methods, data
science, etc., in invited and contributed talks, and less emphasis on update reports.
ADASS talks should be more about the algorithms and techniques than about
the projects themselves.
There was considerable support from the BoF session for invited review
talks looking at the state of play in various aspects of software.

There were several practical suggestions made during the BoF which had the support of the
room: 
Authors should be able to classify their paper under more than one key
theme when submitting an abstract. This would help the LOC when setting up the conference
sessions and ensure that each talk was presented in the most relevant session
from the author's point of view.
There was also a request to allow authors to supply, when submitting
an abstract, an optional link containing more information about the proposed
paper.

Discussion on student attendance saw strong agreement on ensuring that
the conference fee has a student rate, and allowing authors to
indicate whether they are a student both on registration and abstract
submission.
There was also a general feeling that there should be a way to reserve
contributed talk slots for early career attendees (particularly those
who have not previously presented a talk), but the details of how this might be
achieved were not discussed.

Another subject raised was whether some plenary discussion time could
be scheduled during the oral sessions, 
at the end of each major topic, at the end of each
day, or at the end of the conference. There was some support for this and
also some concern, from participants who thought it would cut into the
time available and result in fewer talks. 

A few people showed interested in a hack day, and there was no
opposition to the idea. This could be scheduled as a half-day at the
end of the conference, or be run concurrently with the tutorial.

\item[Posters]
ADASS has a strong tradition of lively and effective poster sessions. The
posters are displayed in a prominent position (close to where coffee
is served) and a good amount of time is allocated to allow people to
view the posters and for meaningful discussions to occur. 
Both the survey results and the BoF discussion expressed support for
maintaining the poster sessions as they are.

A popular suggestion raised at the BoF was to declare two time slots
(one for even-numbered and another for odd-numbered posters)
when people were expected to be in attendance at their posters, to
facilitate discussion.

\item[Birds of a Feather]
BoF sessions at ADASS are an opportunity for
informal, free-flowing discussions among people with a
common interest.
BoFs are scheduled based on proposals submitted by attendees wishing to
organise and chair a session on a given subject.
An average
of 5 BoFs are held each year although the number proposed and held in
any one year varies widely. Some subjects (e.g.\ FITS and other data
formats) have a BoF most years whilst other topics come up once or
twice when they are of interest to the community and then fade away as
their relevance and urgency decreases.

At the XXIII-rd meeting in 2013 informal ``pop-up'' BoFs
were suggested and tried, where the subjects for discussion were
proposed and scheduled during the conference, in the style of an
``unconference''. This was repeated at this year's meeting,
and although the take-up of pop-up BoFs has been small the opinion at the
BoF was that we should continue to offer them.

A suggestion was made to dedicate a session to people's specific
technical problems - ``I've got a problem, does anyone know the
answer?'', harnessing the concentrated expertise of the conference
attendees. It was agreed that pop-up BoFs may be useful here, or
a hack-day, depending on the nature of the problem.

\item[Governance] 
The ADASS Program Organising Committee (POC) consists of around 15
people, these being representatives of the ADASS sponsoring
institutions and members of the community with an interest in
ADASS. Members are nominated to the POC and ratified by the
sitting committee. The 2014 POC
consists of 16 people\footnote{The current POC membership
  can be found at \url{http://adass.org/POC.html}}, and includes an executive committee
comprised of the POC Chair, Vice-chair, Secretary and
Treasurer. 

The survey indicated that the majority of the ADASS community are
satisfied with the conference organisation and governance (67\%) with
smaller fractions preferring the POC to be directly elected from and by the
community (18\%), or wanting to change the length of the term of POC membership (17\%).

It was noted at the BoF session that the POC is gender-unbalanced at
present, and has been so for all of ADASS's history. Additionally the
wish was expressed for the composition of the POC to be more geographically diverse.

\item[Costs and Sponsorship] 
There was agreement at the BoF that we should seek industry
  support for the conference. This was mainly seen as a way to reduce
  costs, and particularly to make it easier for students to
  attend. The practicalities are open to debate, but nobody
  saw it as bad, as long as our independence is preserved.

Once specific idea which saw considerable support was to secure
  sponsorship from a printer company in the form of a printer loan for
  a few days at the start of the meeting.  The posters could then be
  printed at the conference location, avoiding the too-common event of
  posters getting lost or damaged in transit.

The conference fee was discussed, but there was no obvious consensus
on whether lowering the fees would make a difference to people's
ability to attend the meeting, as this is often a fraction of the
travel and accommodation cost.
The comment was made that it is hard for young people to obtain travel
funding to come to ADASS, although it was agreed that it is important
that they do. Student conference rates help slightly as does the
Financial Aid scheme.

Finally, the demonstration fees were deemed to be too high by some survey
respondents, who argued that demo booths are no longer the technological
burden on the LOC that they once were, and that several of the
institutes who wish to have a booth at ADASS are the same ones
who sponsor the conference. 
After discussion the POC agreed to reduce the fee
for demonstrations from research institutions, but
maintain the current fee for commercial bodies.

\end{description}

\section{Where do we go from here?}

The discussion must continue. Please make your ideas and suggestions
known. Speak with a member of the POC, or send your thoughts by email
(\url{poc@adass.org}). Alternatively tweet your suggestions to
(\url{@astroADASS}) or use Appgree to participate in
ADASS polls.

\acknowledgments
On behalf of the ADASS POC the authors would like to thank
all those who took part in the survey, and everyone who 
participated in the BoF discussion.

\end{document}